# All-optical polarization scrambler based on polarization beam splitting with amplified fiber ring

Yuanjie Yu, Shiyun Dai, Qiang Wu, Yu Long, Ai Liu, Peng Cai, Ligang Huang, Lei Gao and Tao Zhu

*Abstract*—Optical-fiber-based polarization scramblers can reduce the impact of polarization sensitive performance of various optical fiber systems. Here, we propose a simple and efficient polarization scrambler based on an all-optical Mach–Zehnder structure by combining polarization beam splitter and amplified fiber ring. To totally decoherence one polarization splitted beam, a fiber ring together with an amplifier are incorporated. The ratio of two orthogonal beams can be controlled by varying the amplification factor, and we observe different evolution trajectories of the output state of polariztaions on Poincaré sphere. When the amplification factor exceeds a certain threshold, the scrambler system exhibits chaotical behavior. A commerical single wavelength laser with linewidth of 3 MHz is utilized to characterize the scrambling performance. We found that when the sampling rate is 1.6 MSa/s, a scrambling speed up to 2000 krad/s can be obtiained for the average degree of polariztion being less than 0.1. We also exploit these chaotic polarization fluctuations to generate random binary number, indicating that the proposed technique is a good candidate for random bit generator.

## I. Introduction

Optical fibers fail to maintain the state of polarization (SOP) of light propagating through them, due to the random birefringence induced by core defects, external stress, bending, temperature, et al [1]. This behavior leads to polarization instability in most fiber systems, which is, in general, a source of additional noise [2]. Polarization scramblers transform the polarized light into non-polarized one to obtain a highly random polarization state, and mitigate the polarization mode dispersion or polarization-dependent loss [3]. The use of polarization scramblers has been reported in many applications such as optical fiber sensors [4,5], fiber-optic gyroscopes [6,7], high precision spectrophotometer [8], fiber amplifiers [9,10], secure key [11], et al. Additionally, they can be used to measure the polarization dependence of fiber-optic components and systems. For that purpose, the scrambling speed induced by the scrambler device should be high enough to match the scale of fast polarization changes encountered in high-speed fiber optic systems [12].

To scramble the polarization, many schemes have been proposed, which have been traditionally enabled by using waveguide electro-optic phase modulation techniques [13-15] or fiber squeezers [16,17]. But the operating speeds of the proposed polarization scramblers are limited by the electronics driving the modulators or the mechanical devices. In the last few years, polarization scramblers based on all-optical interactions have become an alternative and complementary approach to break the speed limitations. For example, an all-optical polarization scrambler structure based on incoherent fiber ring was proposed [18], where input light undergone a series of recirculations inside fiber ring, and they recombined incoherently at the output. Other similar scheme was also demonstrated by cascading the fiber rings [19]. A new structure based on nonlinear interaction between an incident signal and its intense backward replica generated at the fiber-end through an amplified reflective delayed loop was reported [20]. By adjusting the amplification factor $g$, a fully scrambled SOP output and a maximum scrambling speed of 600 krad/s were obtained [21].

Other kinds of polarization scramblers are proposed based on polarization beam splitter (PBS) [22-24]. As depicted in Fig. 1(a), the input light $E(t)$ is splitted into two orthogonal polarization directions as

$$\mathbf{E}(\mathbf{t}) = \begin{bmatrix} \mathbf{E}_x \\ \mathbf{E}_y \end{bmatrix} = \begin{bmatrix} a_x \exp(i\alpha_1) \\ a_y \exp(i\alpha_2) \end{bmatrix} \quad (1)$$

where, $E_x$, $E_y$ are the two orthogonal linear polarized components, $a_x$ and $a_y$ are their amplitudes, $\alpha_1$ and $\alpha_2$ are the corresponding phases. After decoherence completely one orthogonally polarized beam (for example $E_y$) by introducing ultralarge optical delays through additional fibers, they are subsequently recombined. After superposition of the two orthogonal polarized lights, the scramblers exhibit randomness in the polarization space. Recently, we demonstrated a novel structure based on a polarization beam splitting fiber loop, which increased the effective time delay through multiple circulations by fiber loop. The maximum scrambling speed for a chaotic SOP regime reaches 2.5 Mrad/s, at a sampling rate of 1.6 MSa/s [25]. However, we found that the scrambling performance is highly dependent on the SOP of input laser before polarization splitting: the SOP of input laser has to be adjusted precisely so that the amplitudes of two components are equal (state 3 in Fig. 1(a)), or their output SOP may not cover the entire Poincaré sphere. Besides, when the laser coherence is high enough, the required fiber length could be extremely large. The corresponding propagation loss can be intolerable for

This work was supported in part by the National Natural Science Foundation of China under Grant 62075021; in part by the National Science Fund for Distinguished Young Scholars under Grant 61825501 (*Corresponding author: Lei Gao*).

Yuanjie Yu, Shiyun Dai, Qiang Wu, Yu Long, Ai Liu, Peng Cai; Ligang Huang; Lei Gao, and Tao Zhu are with the Key Laboratory of Optoelectronic Technology and Systems (Ministry of Education), Chongqing University, Chongqing 400044, China (e-mail: yuyuanjie@cqu.edu.cn; daisy1@cqu.edu.cn; qiangwu@cqu.edu.cn; longyu@cqu.edu.cn; liuai@cqu.edu.cn; peng_cai@stu.cqu.edu.cn; lghuang@cqu.edu.cn; gaolei@cqu.edu.cn; zhutao@cqu.edu.cn).

practical applications. Most importantly, as shown by Fig.1(b), the extreme large loss in the decoherenced beam makes the amplitude of $E_y$, $a_y$, to be much less than that of $E_x$, namely, $a_x$. Therefore, the recombined output $E'(t)$ exhibits elliptical polarization state, rather than the expected circular polarization state. To be totally random in the polarization space, the amplitude difference between $E_x$ and $E_y$, either induced by propagation loss or polarization splitting, shall be compensated properly (Fig. 1(c)). To be more efficient in practical systems, the delay fiber length also shall be reduced.

In this paper, we propose an all-optical polarization scrambler structure based on the combination of polarization beam splitter and fiber ring (FR). In the component of decoherence, we use an Er-doped fiber amplifier (EDFA) and a fiber ring to replace long single-mode fiber (SMF). The ratio of two orthogonal beams can be controlled by changing the amplification. When the amplification factor exceeds a certain threshold, the system exhibits a chaotic regime in which the evolution of the output polarization state of the signal becomes temporally chaotic and scrambled all over the surface of the Poincaré sphere. We characterize the scrambling performance for 3 MHz linewidth laser. When the sampling rate is 1.6 MSa/s, we obtain a scrambling speed up to 2000 krad/s with an average degree of polarization (DOP) less than 0.1. Moreover, we observe different polarization dynamic evolutions when adjusting amplification factor $g$, indicating that the trajectory for system entering into chaotic regime is determined by the amplitude difference between the decoherenced component and the non-decoherenced one. Then two lasers with different linewidths as input are tested to compare the effect of polarization scrambling. Finally, the chaotic SOP dynamics induced by the scrambler is exploited to generate random binary sequences.

## II. EXPERIMENTAL SETUP

Figure 1(d) denotes the main experimental setup. We used a commercial continuous wave (CW) distributed feedback laser as input source. The polarization controller1 (PC1) is exploited to change the SOP of the input beam in our polarization scrambler. Then, the input signal is split into two orthogonal components by PBS1, namely, $p$- and $s$-wave, which are denoted by double-headed arrows and dots, respectively. Here, one component for $s$-wave is delayed with respect to the other, and then they are subsequently recombined by PBS4. Blue lines denote laser beams propagating in SMF, and red lines indicate those propagating within polarization maintaining fibers. Black lines are electrical signals. The non-polarization-maintaining delayed fiber branch consists of an EDFA followed by a single FR structure as well as a PC2. PBS2, PBS3 are used to connect polarization-maintaining component and SMF. The gain of the EFDA is carefully controlled to adjust the amplification factor $g$. The FR is made of a $2 \times 2$ SMF directional coupler whose coupling ratio is 50:50, and the two pigtails are spliced together to form a loop. PC2 is used to fine-tune the SOP to optimize the output power of this component. The recombination beam is measured by a high-speed polarimeter (Novoptel PM1000) with a maximum sampling rate up to 100 MSa/s. Furthermore, for random bit generation experiments, the output signal SOP is projected on a polarizer (Pol) in order to transfer the polarization chaos into intensity fluctuations. Opto-electrical conversion is fulfilled by photodetectors (PD) with analog bandwidth of 8 GHz, and resulting random signals are acquired by high-speed oscilloscope (Tektronix, DSA72004B) with bandwidth of 20 GHz and a sampling rate of 50 GSa s-1. The degree of randomness of the obtained data is analyzed quantitatively by the autocorrelation and cross-correlation method.

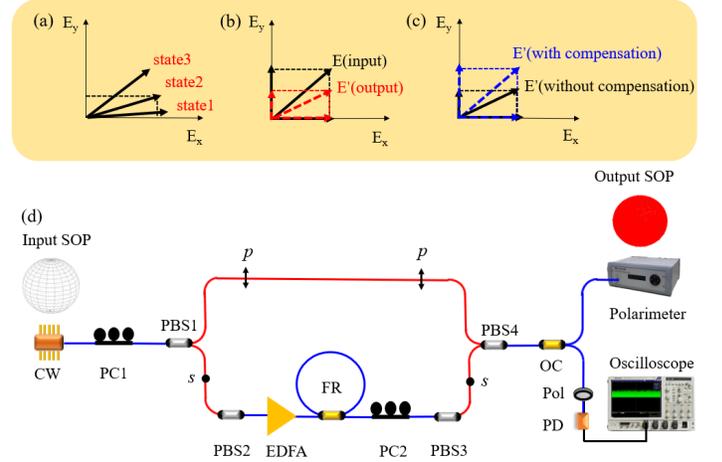

**Fig. 1.** (a) Amplitude difference between $E_x$ and $E_y$ for input lasers with three different polarization states, whose polarizations are split by a PBS. (b) Amplitude difference between $E_x$ and $E_y$ induced by propagation loss, even the two amplitudes are equally split by a PBS in the input E (input). (c) Amplitudes of two components are compensated to be equal. The original inequality is induced either by propagation loss or polarization splitting. (d) Experimental setup of the polarization scrambler. PBS: polarization beam splitter, FR: fiber ring, PC: polarization controller, Pol: polarizer, CW: continuous wave, PD: photodetector, OC: optical coupler.

## III. RESULTS AND DISCUSSION

For a polarization scrambler based on a Mach–Zehnder fiber delay-line structure, to reduce the mutual coherence of the orthogonal polarization modes and realize polarization scrambling effectively [22], the length of the fiber delay line must be longer than the coherence length of the light source. For a given CW laser with linewidth of $\Delta v$, the corresponding reference length $L$ in fiber is given by [26]:

$$L = \frac{L'}{n} = \frac{c}{n\Delta v} \quad (2)$$

where, $L'$, $c$, and $n$ are the coherence length, the speed of light, and the fiber refractive index, respectively. The $2 \times 2$ fused single-mode coupler with 1 m of fiber pigtail at each port yields a delay line ~2 m in length. Once the group delay induced by propagation of recirculating length is larger than the coherence time of the laser source, the mutual coherence of the orthogonal polarization components can be totally depressed. Then, as each recirculating beam passing through the FR is incoherent with others, all the beams recombine incoherently at the output, and the DOP of the output light is reduced drastically [18].

## A. Polarization scrambling comparisons for different polarization states

To be more representative and without losing generality, we set the amplification factor *g* to be 1, namely the amplification would not influence the polarization scrambling performance. The SOPs of input laser are adjusted to be three states, as indcted schematically in Fig. 1(a). The corresponding outputs on the Poincaré sphere for the three input states are depicted in Fig. 2 (a). For state 1, the energy of the input lasers after polarization splitting via PBS1 is mainly contributes to the *p*-wave, and the corresponding output SOP is concentrated in a small area. Here, the decoherence in the *s*-wave branch do not contribute significantly to the combined output after superposition via PBS4. It is obvious that when the SOP of the input lasers are adjusted from state 1 to state 2, through properly setting PC1, the amplitudes for the two polarized components are gradually to be comparable, and the output SOPs exhibit more randomly on the Poincaré sphere. The reason is clear: the larger the intensity of *s*-wave branch, the more contribution the decoherence to the combined lasers. An optimal polarization scrambling performance occurs for the state 3, where the amplitudes of the two polarized components are almost equal.

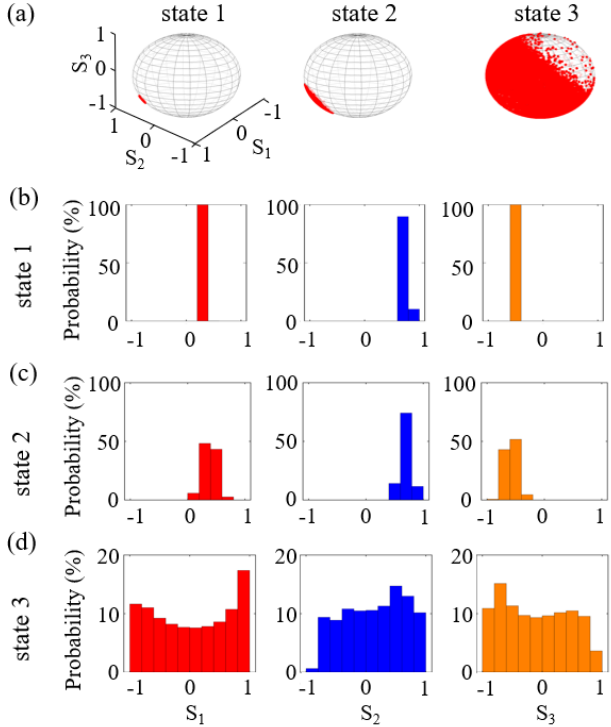

**Fig. 2.** (a) Three typical outputs for different input states selected by adjusting PC1. All Poincaré spheres are with the same coordinate frame and only first one is shown for a better display. (b)-(d) Probability distribution histograms of the three normalized Stokes parameters $S_1$, $S_2$ and $S_3$.

The polarization scrambling performance can be quantitatively revealed by the probability distributions. We characterize the experimental probability histograms of three normalized Stokes parameters $S_1$, $S_2$ and $S_3$ for each state. The points on the Poincaré sphere in state 1 are concentrated around a fixed point, and the corresponding probability histogram is located at a certain bin of the Stokes parameters (Fig. 2(b)). As the points scatter to a small area in state 2, the probability density function (PDF) is unevenly distributed between -1 and 1 (Fig. 2(c)). In state 3, the PDF of all normalized Stokes parameters appears almost uniform (Fig. 2(d)), which provides further evidence of the polarization randomization. Note that a perfectly uniform density function of the Stokes components would correspond to the ideal chaotic regime.

## B. Influence of amplification factor on the polarization scrambling performance

To fully scramble the SOP, the amplitude difference between *p*- and *s*-wave caused by polarization splitting or propagation loss shall be compensated properly. In our configuration, field intensities of the two superimposed components in PBS4 can be manipulated to be equal by properly setting amplification factor *g*. Here, we investigate the polarization scrambling performance for different amplification factors. The input state 1 is taken for example, and similar evolution dynamics can be found for other states. Figure 3 depicts the evolution trajectories of the output SOPs on Poincaré sphere for state 1 when increasing the amplification factor *g*. The obtained output SOPs are projected into the $S_1$-$S_3$ plane. Three different regions for various *g* parameter can be observed, ranging from an ordered pattern to a chaotic pattern and then to a partial chaotic pattern. By finely tuning PC1, when *p*-wave energy predominates, the output pattern is a single point on the Poincaré sphere, corresponding to the ordered state, with the amplification factor *g* being 1. In this case, due to fixed PC1 after adjustment, the scrambling process is relative weak. When the amplification factor *g* increases slowly, the system becomes unstable and begins to oscillate. Further improving the power of EDFA, polarization points spread gradually. For *g* ranging from 250 to 350, where the intensities of the two components are approximately equal, the trajectory of output SOP almost uniformly covers the complete surface of the sphere, achieving nondeterministic and totally random polarization scrambling of the output signal. The corresponding DOP is about 0.08. By continuingly increasing *g*, the power of *s*-wave predominates of the recombined laser, and the polarization distribution of output lasers evolve from one semisphere to the opposite semisphere. The corresponding DOP increases to 0.54, and output SOP becoming partially chaotic.

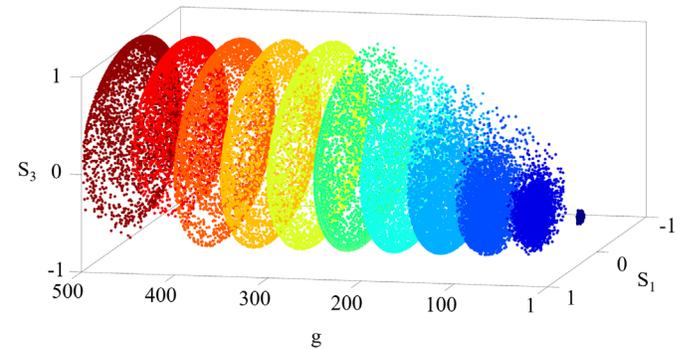

**Fig. 3.** Projection of the experimentally measured output SOPs of state 1 in the $S_1$-$S_3$ plane.

The scrambling performances can be evaluated quantitatively by scrambling speed *V* and the averaged DOP, which are defined as follows [21]:

$$V = \left\langle 2\arcsin\left(\frac{1}{2}\frac{\partial \vec{s}(t)}{\partial t}\right)\right\rangle \quad (3)$$

$$\mathrm{DOP} = \frac{\sqrt{\langle S_1(t)\rangle^2 + \langle S_2(t)\rangle^2 + \langle S_3(t)\rangle^2}}{\langle S_0(t)\rangle} \quad (4)$$

where, $t$ represents sampling time, $S_0$ denotes total intensity, $S_1$, $S_2$ and $S_3$ are the normalized Stokes parameters. To investigate the polarization scrambling performance for different amplification factors, we carried out measurements of polarization scrambling speed and DOP for the three different initial states above. To show the SOP scrambling evolution process in detail, the sampling rate of the polarimeter is set as 1.6 MSa/s. Figures 4(a)-(c) summarize the experimental results about the output DOP and corresponding scrambling speed $V$ as a function of $g$ for the three input states.

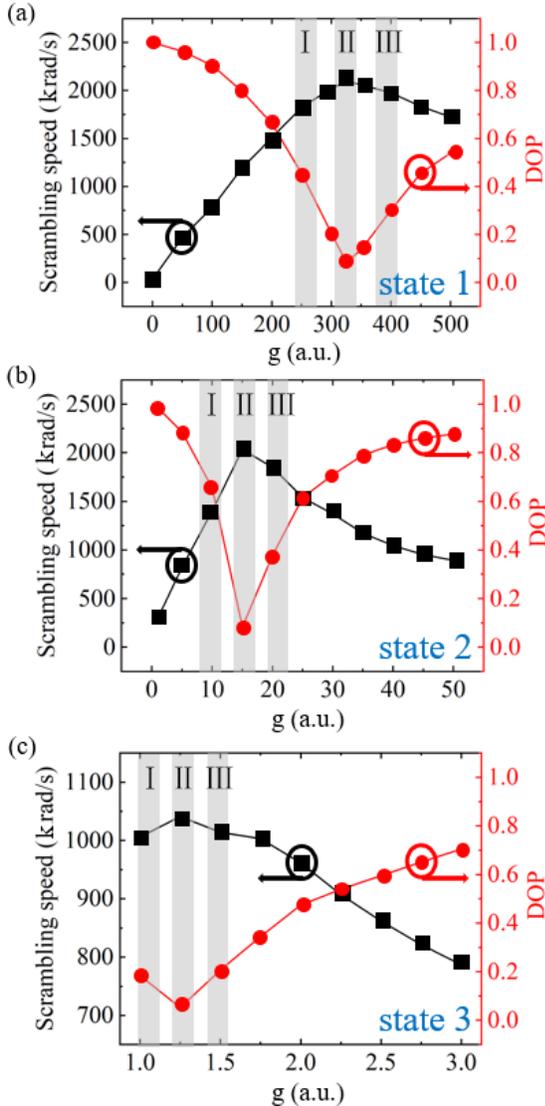

Fig. 4. Polarization scrambling performance for state 1 (a), state 2 (b), and state 3 (c). The black square represents $V$, and the red dot represents DOP.

For both the three input states, an optimized amplification factor can be found for a fully random SOP output, which can be inferred by the DOP less than 0.1. While, the required amplification factors for the three states are totally different. For state 1, owing to the dominance of $p$-wave energy, a much larger amplification factor $g$, is needed so that the amplified $s$-wave is comparable to the $p$-wave. As depicted in Figs. 4(a)-(c), as the amplification factor $g$ increases steadily, the scrambling speed also increases continuously, whereas the DOP drops from 1 to almost 0. When the polarization scrambler is fully scrambled, $V$ reaches its maximum value and DOP becomes less than 0.1. With further increasing $g$, the amplified $s$-wave is much larger than the $p$-wave, and the corresponding $V$ and DOP changes in the opposite direction. The obtained two lines are consistent with the SOP evolution on Poincaré sphere illustrated in Fig. 3. Similar trends are obtained for the state 2 and state 3. However, we found that the maximum scrambling speed (state 1 and state 2) is around 2000 krad/s, whereas it is about 1000 krad/s in state 3. The inconsistency of scrambling speed can be attributed to spontaneous emission noise of amplifier. When the amplification factor is low, the weak spontaneous emission has little impact on polarization. However, as the amplification increases, spontaneous emission facilitates the acceleration of polarization scrambling.

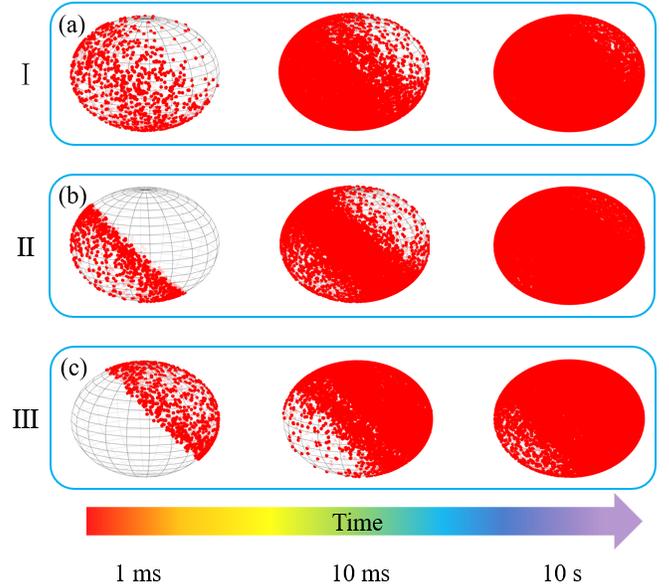

Fig. 5. (a)-(c) The trajectories of SOP entering into chaotic regime in region I, II, III, for state 1, respectively.

In addition, our observations indicate that the trajectory of system entering into chaotic regime is different when adjusting the amplification factor $g$. Specifically, in Figs. 4(a)–(c), we have identified three distinct regions, I, II, and III, each displaying unique dynamics. Here, we also take the state 1 as an example, and the SOP trajectories for different amplification factors within 10 s are depicted in Fig. 5. For region I (Fig. 5(a)), where the power of $p$-wave is dominant, the SOP is initially distributed on the left, and it gradually spreads to the right over time until it covers the entire sphere. Within region II (Fig. 5(b)), the field intensities of two superimposed components are equal, and the SOP forms a loop in the middle of the sphere before rapidly extending to both sides. As $g$ is increased further, the system transitions to region III, where the power of $s$-wave is greater than that of $p$-wave. The evolution track of SOP transits from right to left over time until it covers almost homogeneously on the surface of the Poincaré sphere (Fig. 5(c)). This trajectory is opposite to the behavior in region I. More details about the polarization trajectories are visualized by the movies in supplementary file.

*C. Polarization scrambling performance for lasers with different linewidths*

The proposed polarization scrambler structure can be utilized for lasers with any linewidths as long as the delay fiber in the FR is properly chosen. Here, given the specific parameters in our proposed structure, we compare the polarization scrambling performance for two input lasers with totally different linewidths, one with 3 MHz, while the other is 200 kHz. As shown in Figs. 6(a) and (d), if the SOP entering into PBS1 was mainly aligned on *p*-wave in the initial state, the output SOP is concentrated in a small area. In order to fully scramble the SOP, it is necessary to increase the power of *s*-wave, through which the polarization scatters more intensely on the Poincaré sphere (Figs. 6 (b) and (e)). Finally, an optimized amplification factor can always be found, so that the amplitude of the two orthogonal components are equal, and the optimized polarization distributions can be found in Figs. 6 (c) and (f), where both SOPs fluctuate with largest areas on the Poincaré sphere. Yet, performance difference is obvious as the two lasers have different linewidths. For laser with linewidth of 200 kHz, as the *s*-wave branch cannot be decoherenced completely by the FR, the evolution of the track only covers along a ring with certain thickness, while the lower left and upper right corners still cannot be filled (Fig. 6(c)). However, for the laser with linewidth of 3 MHz, decoherence can be achieved completely, and the SOPs spread from a fixed area, to a semi-circle, and finally to be random on the full Poincaré sphere (Figs. 6(d)-(f)).

$\Delta v = 200\ kHz$

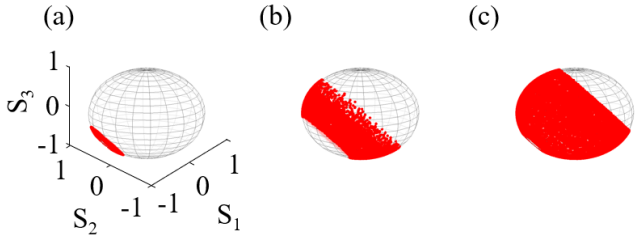

$\Delta v = 3\ MHz$

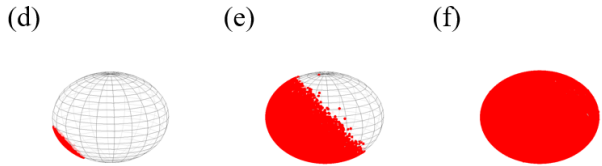

**Fig. 6.** (a)-(c) Polarization scrambling states of a 200 kHz laser under different amplification factors. (d)-(f) Polarization scrambling states of a 3 MHz laser under different amplification factors.

*D. Polarization scrambler for random bit generation*

Polarization chaos in optical fiber can be an efficient source of randomness for the generation of random numbers[21,27,28]. Here, we demonstrate the practical application of our polarization scrambler as source for random binary sequences. The fluctuations of the output light intensity are exploited to experimentally generate random sequences. The principle of operation of random bit generation from polarization chaos is to convert the output field intensities of the scrambler into either a 0 or 1 according to some specific threshold, here calculated from the median value of the waveform. According to the setup in Fig. 1(d), after the projection on a polarizer, polarization randomness of the scramblered signal is transformed into the temporal fluctuations. Figure 7(a) displays a part of one experimental raw data (blue line) as well as clock signal (red line). Our clock has been chosen to 250 kHz whose frequency is selected below the typical correlation length of the test signal so as to ensure a reliable randomization [21]. The experimental raw data is sampled at each rising edge of the clock. After setting the threshold to the median value of 0.1618, the binary random sequence represented in Fig. 7(b) has been obtained.

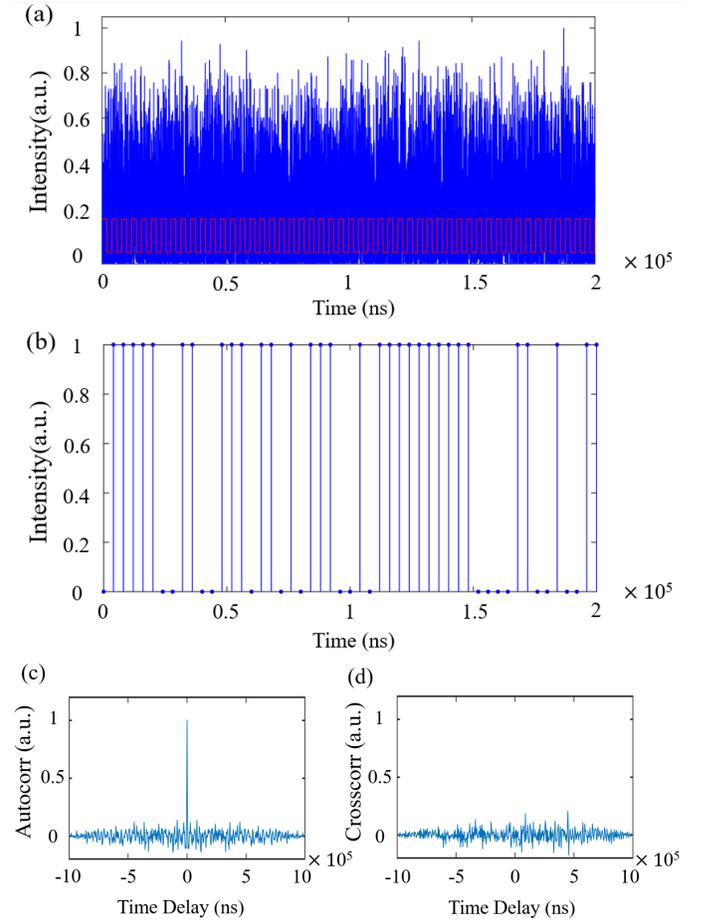

**Fig. 7.** (a) Evolution of the light intensity over time (blue line). Square signal taken in correspondence of a clock (red line). (b) Random sequence generated after thresholding samples chosen in (a). (c) Autocorrelation of random sequence in (a). (d) Cross-correlation function of different random sequences generated in experiments.

The degree of randomness of the obtained temporally random sequence are tested through calculation of correlation function [21]. Since the relative average value of temporal intensity is 0.5, we normalize it before the correlation calculations. The autocorrelation function for the zero-mean normalized binary sequence in Fig. 7(a) is depicted in Fig. 7(c), and the cross-correlation function between two different sequences is shown in Fig. 7(d). Both results show that sequences are autocorrelated only when there is a delay of 0, and and there is no significant correlation between two different sequences. All sequences generated through polarization chaos are random, which would be an effective method for generating random bit.

IV. CONCLUSION

In summary, we have proposed an all-optical polarization scrambler structure by combination of polarization beam splitter and fiber ring. To compensate the amplitude difference induced by propagation loss or polarization splitting, optical amplification provided by EDFA are utilized in the decoherence component, for fully scrambling SOP of continuous wave lasers. By controlling the amplitude difference of the two orthogonal beams through changing the amplification factor $g$, we observed different polarization evolution trajectories for the system entering into chaotic regime. For such a scrambler, we obtained a scrambling speed up to 2000 krad/s for a 3 MHz linewidth laser, where the DOP is less than 0.1. We also described how the chaotic SOP fluctuations induced by the scrambler can be exploited to generate ensembles of random bit, by converting the polarization chaos of the output signal into temporal intensity fluctuation, which can generate random binary sequences by a digitalization process. The proposed polarization scrambler can reduce the impact of polarization sensitivity for various optical sensing systems, and is also reliable as random signal source for communications, optical signal process, et al.